# Reconstructing the Star Knowledge of Aboriginal Tasmanians

Michelle Gantevoort

Nura Gili Indigenous Programs Unit, University of New South Wales, Sydney, NSW, 2052, Australia
Email: gantevoort@icloud.com

Duane W. Hamacher

Monash Indigenous Centre, Monash University, Clayton, VIC, 3800, Australia.
Email: duane.hamacher@monash.edu

Savannah Lischick

LifeCell Corporation, 5 Millenium Way, Branchburg NJ, 08876, United States.
Email: Savannah.lischick@gmail.com

**Abstract**

The canopy of stars is a central presence in the daily and spiritual lives of Aboriginal Tasmanians. With the arrival of European colonists, Tasmanian astronomical knowledge and traditions were interrupted and dispersed. Fragments can be found scattered in the ethnographic and historical record throughout the nineteenth century. We draw from ethno-historical documents to analyse and reconstruct Aboriginal astronomical knowledge in Tasmania. This analysis demonstrates that stars, the Milky Way, constellations, dark nebula, the Sun, Moon, meteors, and aurorae held cultural, spiritual, and subsistence significance within the Aboriginal cultures of Tasmania. We move beyond a monolithic view of Aboriginal astronomical knowledge in Tasmania, commonly portrayed in previous research, to lay the groundwork for future ethnographic and archaeological fieldwork with Aboriginal elders and communities.

**Warning to Australian Aboriginal readers:** This paper contains the images of Aboriginal people who have died.

**Keywords**: Cultural Astronomy; Ethnoastronomy; Indigenous Knowledge Systems, Aboriginal Australians, Tasmania

"*Aboriginal Tasmanians spoke of the subject of stars with great zest.*"
- George Augustus Robinson, 13 March 1834.

## 1 Introduction

The study of Indigenous Knowledge Systems can reveal a wealth of information about how scientific information is encoded in oral tradition and material culture (Agrawal, 1995), particularly with respect to astronomical knowledge (Cairns and Harney, 2003; Hamacher, 2012; Fuller et al., 2014; Norris, 2016). The continued study of Aboriginal and Torres Strait Islander astronomical knowledge, and the traditions though which this knowledge is passed to successive generations, has led to a more detailed understanding of how the sun, moon, and stars aided navigation, seasonal calendars, food economics, animal behaviour, social structure, sacred law, and relationships between the land and the sky (Johnson, 1998). This is done through the various methodologies and theoretical frameworks of cultural astronomy, an in-terdisciplinary academic field that seeks to understand the role and use of astronomy in culture (Ruggles, 2015).

Ethnohistorical literature is one of the primary sources for studying and reconstructing Indigenous astronomical knowledge (Hamacher, 2012). Aboriginal Australians are considered to be among the oldest continuous cultures, and *the* most researched Indigenous people, on Earth (Smith, 1999: 3), with records of language, customs, and traditions going back to before first colonisation in 1788. However, these records are highly biased, as Aboriginal people were considered to be among the lowest rung of human cultures by the colonists. This false position, and the rapid decimation of Aboriginal people and culture after British colonisation, lead to the practice of "salvage anthropology", where ethnographers sought to record Aboriginal traditions before the people and cultures "disappeared." This lead to a rather large body of published information about Aboriginal cultures. Unfortunately, much of the astronomical knowledge from these records is highly fragmented and incomplete. A lack of formal training or understanding of astronomy by these ethnographers means much of the recorded information is filled with conflated terminology, misidentifications, incorrect assumptions, and transcription errors.

Aboriginal Tasmania has long been a place of contrasts, contention, and devastation (Ryan, 1996). Colonialism, dispossession, genocide, and disease nearly wiped out Tasmanian Aboriginal people. Before the arrival of Europeans, it is believed Aboriginal people arrived in Tasmania (Troweena) over 40,000 years ago (Pope and Terrell, 2007; Lourandos, 1997) when the region was connected to mainland Australia by a land bridge (Orchiston, 1979a,b; Murray-Wallace, 2002). Approximately 8,000 years ago, rising sea levels created the island of Tasmania, separating Aboriginal Tasmanians from mainland Aboriginal people. It was believed that the Tasmanian Aboriginal people remained relatively isolated until European contact and subsequent colonisation (Johnson et al., 2015: 16). Aboriginal groups are spread across roughly nine territories (*ibid*: 36): Northeast, Ben Lomond, North Midlands, Oyster Bay, Southeast, Big River, North, Northwest, and Southwest (Figure 1). Within each of these territories existed smaller groups tied through marriage, kinship, and language, led by a respected male elder (*ibid*).

In 1996, there was a split among Aboriginal Tasmanians into two major groups: the Palawa and the Lia Pootah. The Palawa take their name from the first man created from a kangaroo by a creator spirit. The Palawa claim to be descendants of two people: (1) Dolly Dalrymple (c.1808-1864), an Aboriginal woman born on the Furneaux Islands in Bass Strait of European sealer George Briggs and Aboriginal woman, Woretemoeteyenner (who was abducted by Briggs; McFarlane, 2005), and (2) Fanny Cochrane Smith (1834-1905), an Aboriginal Tasmanian woman believed to be the last fluent speaker of a Tasmanian language, who was born on Flinders Island to Aboriginal Tasmanian parents (Clark, 1988). The Lia Pootah claim that not all Aboriginal Tasmanians were removed from the island or died-off, stating that they are descendants of the Aboriginal women of the Big Rivers and Huon regions of Tasmania and European men from the early days of colonisation (1803 onwards). Some Lia Pootah also claim ancestry from Bruny Island and regions on the east coast and central highlands (Anonymous, 1996). Contention surrounds government and Palawa recognition of the Lia Pootah (e.g. Anonymous 2000; Denholm, 2007). We utilise the term "Palawa" in this paper, but this is not reflection or position on the debate between these two groups.

Relatively little is known about Palawa cultures prior to colonisation. Ethnographic studies were limited and the focus of colonial presence in Tasmania became one of complete Aborig-inal removal from the island. Following a series of conflicts between colonists and Aboriginal Tasmanians in the early nineteenth century - a period known as the Black War - the builder and religious man George Augustus Robinson was hired from 1829-1834 to find the remaining Palawa living in Tasmania, facilitate their "peaceful surrender", then relocate them to Flinders Island. This "Friendly Mission" was accomplished by 1835 and many of the 200 Palawa relocated died from poor health and the prison-like conditions in which they were held. This had a devastating impact, resulting in the near decimation of Palawa culture, traditions, and languages. In the time since, a cultural revival has taken hold and a resurgence of Palawa language, archaeology, history, and culture is rapidly growing.

Because of colonisation, disease, dispossession, and genocide, we know relatively little about Palawa astronomical knowledge. Most of the archival information is ethnohistoric in nature, having been recorded by colonists and missionaries from their Aboriginal contacts, and much of that is fragmented, incomplete, sometimes ambiguous, and always recorded through the lens of the coloniser. Some of the traditional knowledge has survived with the Aboriginal people, who continue to pass their traditions to successive generations.

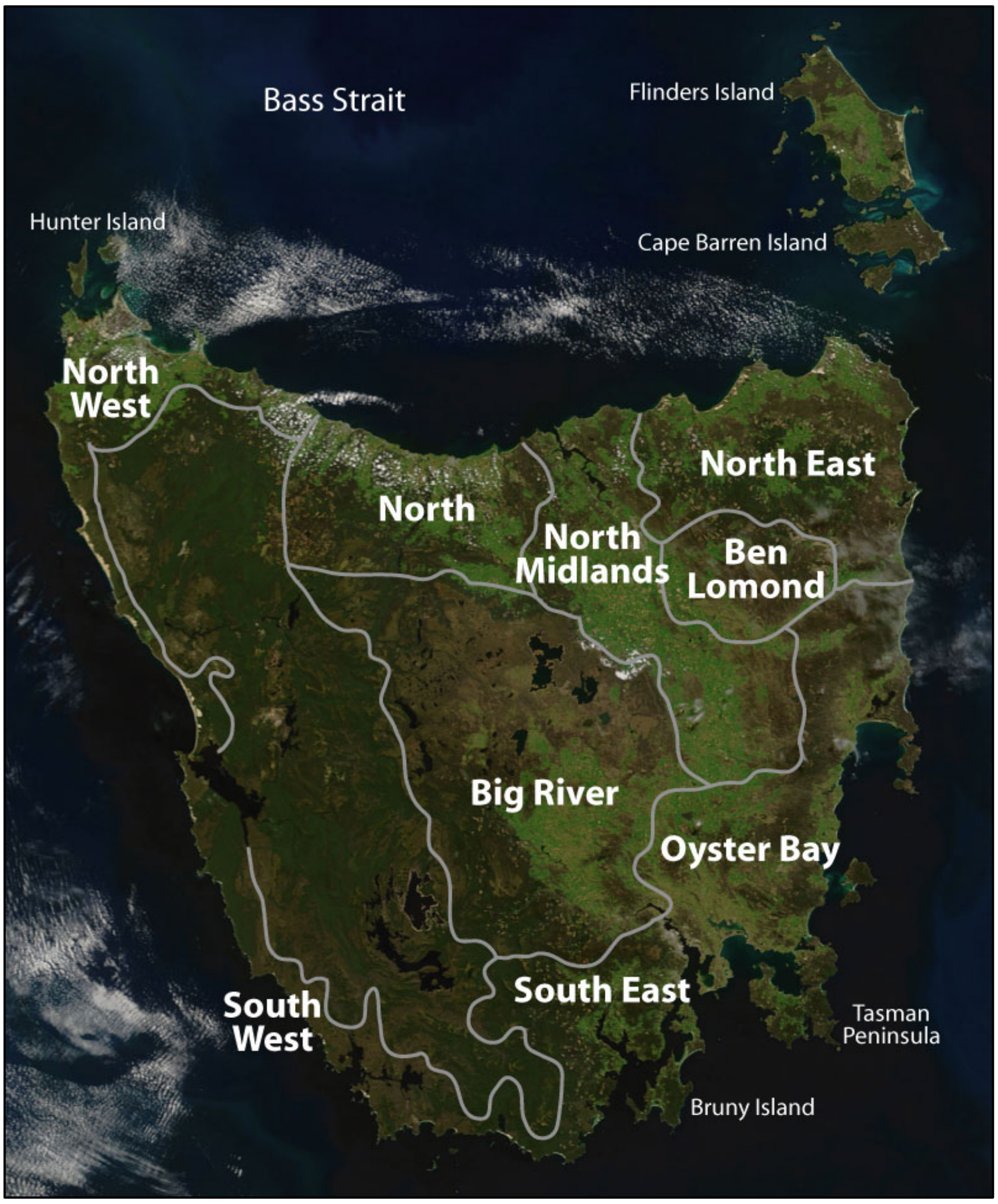

*Figure 1: Map of major Tasmanian groups at the time of first European contact. Image: Wikimedia Commons License.*

This paper attempts to sort though the fragments of astronomical knowledge from Aboriginal Tasmanians scattered throughout the ethnohistorical literature and archives, re-analyse them using established and emerging methodologies from cultural astronomy, and attempt to reconstruct this knowledge to the best of our ability, though we acknowledge our limitations in this endeavour. This will serve as a base for further ethnographic and archaeological studies, with application to education and cultural revival.

## 2 Methodology

This research draws upon the ethnohistoric documents and published material in the literature, newspapers, library and museum archives, and any associated media that makes any mention of astronomical objects or phenomena with respect to Palawa traditions. No ethnographic fieldwork was conducted for this project.

We identify the original Aboriginal sources of information, linking the accounts, narratives, or descriptions whenever possible. Two key sources of information are *Mannalargenna*, a leader of the northeast Palawa, and *Woorrady*, a Nueone man of Bruny Island/Lunawanna-Alonnah. They guided Robinson through Tasmania in the 1830's. As they spoke of their people's culture, Robinson recorded in his journals. Much of the cultural knowledge was recorded during a period of rapid growth of the colony. As such, many of Robinson's (and others') records do not name the Aboriginal sources of these oral traditions. Sometimes only the region from which the tradition was recorded is provided.

The Palawa in Table 1/Figure 2 are potential sources of astronomical knowledge despite not being specifically identified in the written record. They were members of Robinson's expedition team, guides, or close relations to those guides. In addition to Robinson, Palawa information was recorded by Henry L. Roth (1899), James B. Walker, Joseph Milligan (1890), and James Bonwick (1870; 1884). These men recorded features of Palawa astronomy, usually within a broader discussion of Western ideas of religion and spirituality. Robinson's journals are frequently regarded as the most detailed written account of Palawa life available. During Robinson's mission from 1829 to 1834, he documented his interactions with Aboriginal people in his journals, which were later published as *The Friendly Mission* (Robinson and Plomley, 2008). It is within these journals that we find a majority of the references to Palawa astronomical knowledge.

*Table 1: Palawa who accompanied Robinson on his mission and expedition who are likely sources of the recorded in formation in Robinson's journals. Timler was not part of Robinson's expedition, but is cited as a source of oral traditions by Cotton (1979).*

| Name | Life | Description | Reference |
|---|---|---|---|
| Bullrer | ca 1812-1845< | Pairrebeenne woman from Tebrikunna in the far northeast of Tasmania. She joined Robinson's expedition in 1830 | http://www.utas.edu.au/telling-places-in-country/historical-context/historical-biographies/bullrer |
| Calamarowenye | ca 1812-1860 | A man from the Big River region, also called Kalamaruwinya, and later husband of Bullrer. Participated in guerrilla attacks against the colonists during the Black War. He kept the jawbone of his murdered brother as a protective amulet, but it was taken by Robinson. | http://tacinc.com.au/wp-content/uploads/2015/11/MANGALORE-PONT_ABORIGINAL_VALUES.pdf |
| Kickerterpoller | ca 1803-1832 | Paredarererme man, also known as Kikatapula, was kidnapped by colonists at age 9 and broke free in 1822 and joined the Aboriginal resistance against the colonists. He joined the Robinson expedition because of his multi-lingual abilities. | http://www.utas.edu.au/telling-places-in-country/historical-context/historical-biographies/kickerterpoller |
| Mannalargenna | ca 1775-1835 | Leader of the Pairrebeenne clan (Cape Portland) in northeast Tasmania. He lead guerrilla attacks against the colonists during the Black War and was part of Robinson's team. It is believed that his secret intentions were to lead Robinson away from the people Robinson was trying to find and relocate. | http://www.utas.edu.au/telling-places-in-country/historical-context/historical-biographies/mannalargenna |
| Marlapowaynererner | ca 1825-1842 | Son of a clan leader Raleleeper from Georges Rocks who joined the Robinson expedition in 1830, serving to 1835. Also known as Timme, he joined Tunnerminnerwait in Victoria and was hanged alongside him in 1842. | http://www.utas.edu.au/telling-places-in-country/historical-context/historical-biographies/timme |
| Tanleboneyer | ca 1807-1835 | Mannalargenna's second wife and one of the guides on the Robinson expedition who joined in 1830. | http://www.utas.edu.au/telling-places-in-country/historical-context/historical-biographies/tanleboneyer-sall |
| Timler | Unknown. Alive in 1830s | Timler was an elder of the Big River people who recounted some of his stories to Joseph and Isobel Cotton in the 1830s. He was not a member of Robinson's expedition. | Cotton (1979) |
| Truganini | ca 1812-1876 | Woorrady's young wife and daughter of Mangana, the Bruny Island leader. | http://www.utas.edu.au/library/companion_to_tasmanian_history/T/Truganini.htm |
| Tuererningher | ? - 1837 | Bruny Island woman and sister to a female Nuenonne clan leader known as Nelson. Joined Robinson expedition in 1829 through 1835, before dying in 1837. Kickerterpoller was her second husband (her first was Mangana, who died in 1829). | http://www.utas.edu.au/telling-places-in-country/historical-context/historical-biographies/pagerly |
| Tunnerminnerwait | ca 1812-1842 | Pairelehoinner man from Cape Grim, also known as Peevay. Joined Robinson expedition in 1830 and served until 1835. Later charged with killing two whalers in Victoria and hanged on 20 January 1842. | http://www.utas.edu.au/telling-places-in-country/historical-context/historical-biographies/peevay |
| Wapperty | ca 1797-1867 | A Pairrebeenne woman and daughter of Manalargenna (one of four). Many Palawa today are descendants of Wapperty. | Lydon (2014: 37) |
| Woorrady | ca 1784-1842 | Nueone man of Bruny Island, also known as Mutteellee , who was one of Robinson's guides. he joined the team when he was about 45 years old, showing concern about approaching other clansmen with the team. He died of "senility" at age 58. | http://www.utas.edu.au/telling-places-in-country/historical-context/historical-biographies/woorrady |

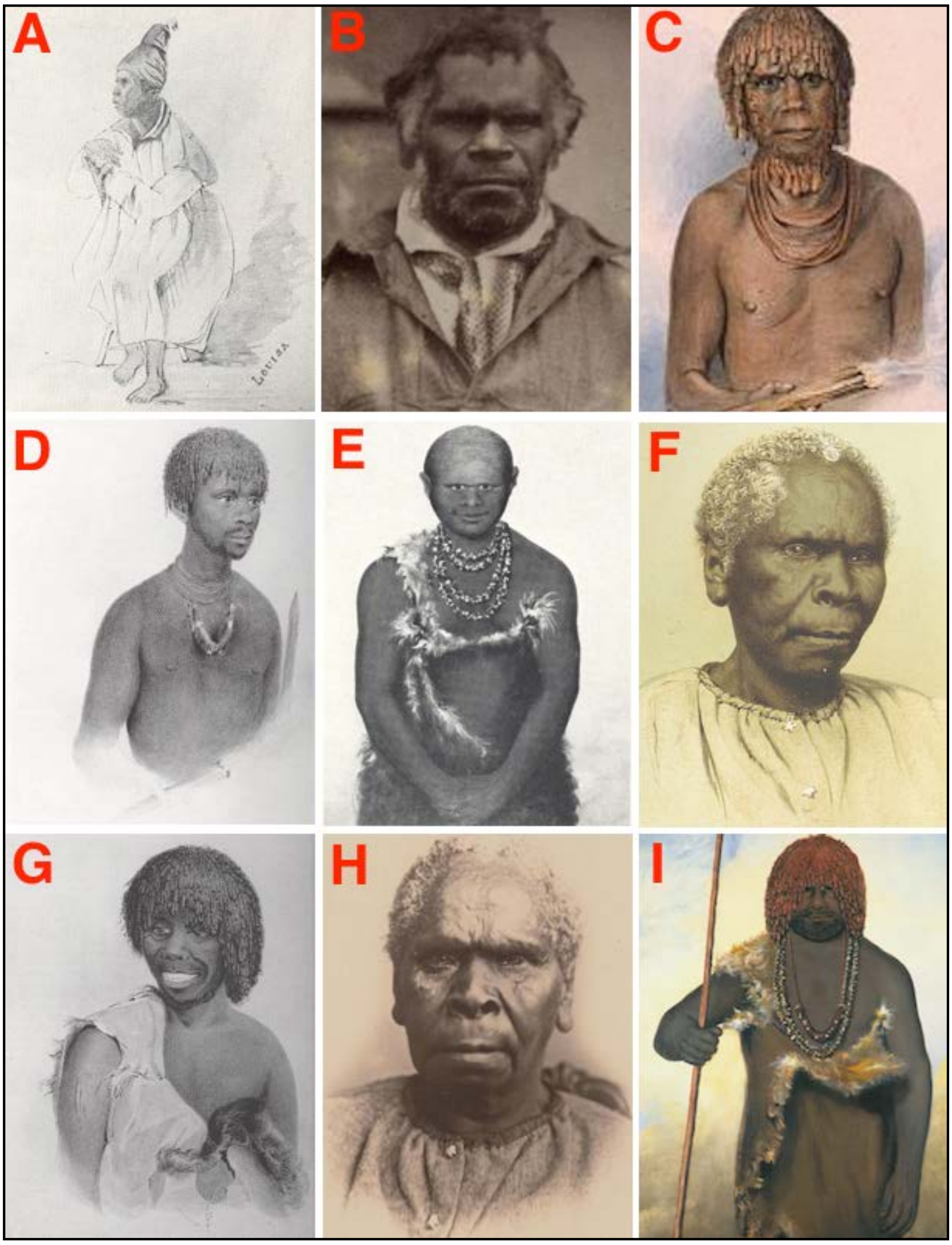

*Figure 2: Aboriginal people who were, or may have been, sources of knowledge for Robinson (listed alphabetically): (A) Bullrer, (B) Calamarowenye (State Library of New South Wales), (C) Mannalargenna (British Museum), (D) Marlapowaynererner, (E) Tanleboneyer, (F) Truganini (Cavendish and Cavendish, 1871: 187), (G) Tunnerminnerwait, (H) Wapperty (National Library of Australia), (I) Woorrady (Tasmanian Museum and Art Gallery). Images (A), (D), (E), and (G) taken from the "Telling Places in Country" site at the University of Tasmania (www.utas.edu.au/telling-places-in-country/historical-context/historical-biographies)*

One of the problems with examining colonial records is that they are translated into Western terminology by non-Aboriginal recorders, who often have a very limited understanding of the Aboriginal traditions. This is further complicated if the recorder does not have a detailed knowledge of astronomy. Misidentifications, conflated terminology, and transcription errors plague colonial records of Aboriginal astronomical knowledge (e.g. Hamacher, 2012; Leaman and Hamacher, 2014). Limited information is provided about the identities of the stars in Palawa traditions, and some seem inconsistent or unlikely. In this paper, we will examine the identification proposed by other researchers, and then offer those we think best fits the information provided. This is aimed at obtaining the best picture of Palawa astronomical traditions in the most rigorous way possible. This will form the basis of future work with Palawa elders.

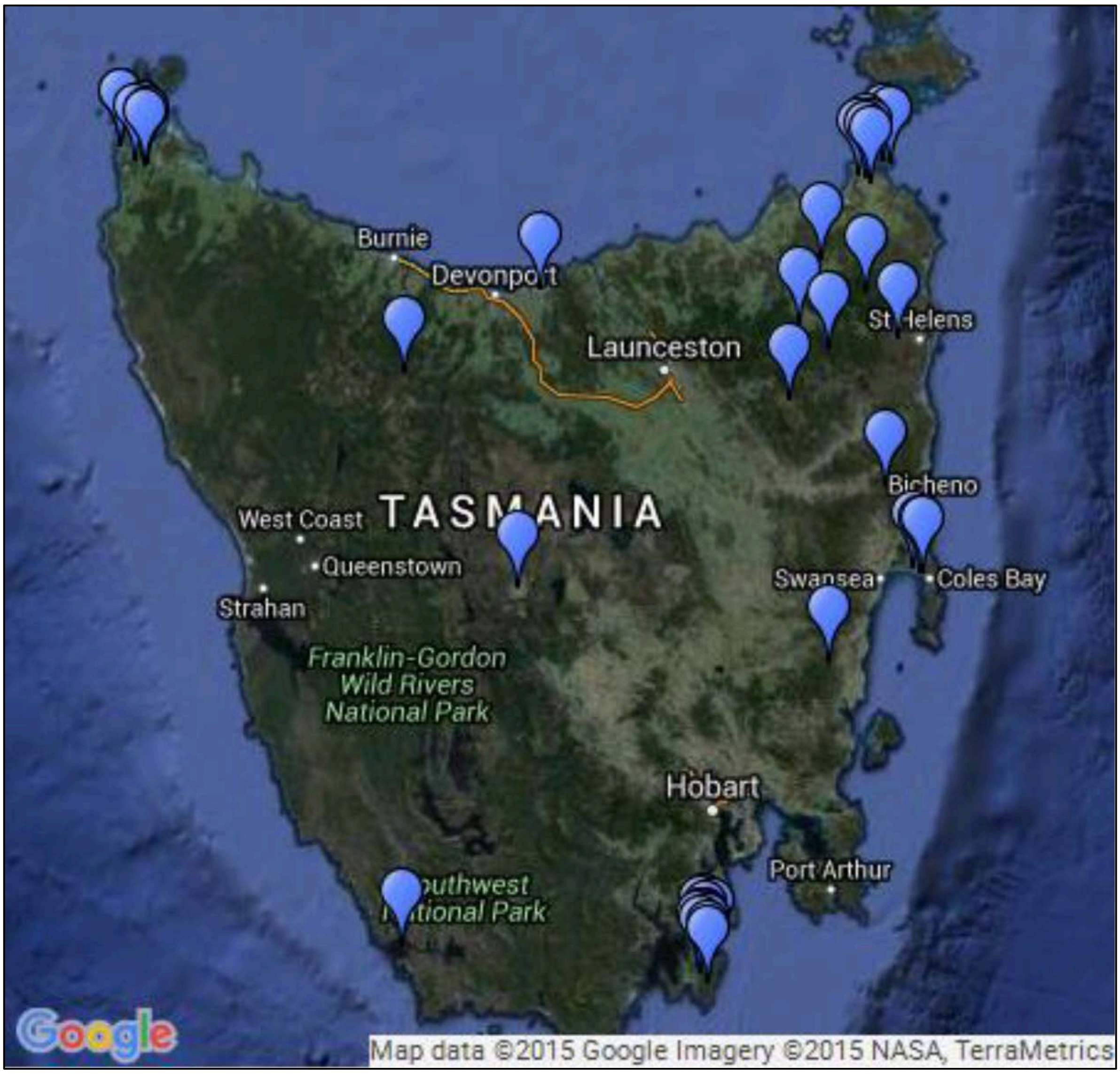

*Figure 3: A map of the locations Palawa astronomical information was taken, using the journals of Robinson and other sources. Image from Google Maps.*

## 3 RESULTS AND ANALYSIS

A review of the archival documentation reveals 42 accounts of Aboriginal astronomy tied to a physical location in Tasmania. These locations are pinned on the map in Figure 3. The astronomical traditions include stars, constellations, and celestial objects (14), the Moon (5), The Sun (1), planets (2), and ancestral spirits connected to the stars (8). These are divided between Bruny Island (13), the Northwest (8), Cape Portland and Swan Island (7), Oyster Bay (5), the Northeast (5), Port Sorell (1), the Big River (1), and Ben Lomond (1).

We present the results in the following themes (A) ***Cosmogony***: Palawa traditions that describe the formation of the land and the creation of the first people; (B) ***Stingray in the Sky***: a Palawa constellation, attempting to identify the celestial objects involved in the tradition, as their Western counterparts are not explicitly named; (C) ***Time and Astronomy***: different concepts of time and the ways astronomical objects were used to denote time reckoning and seasonal change; (D) ***Lunar traditions***: Palawa views of the moon; and (E) ***Transient Phenomena***: Palawa traditions of aurorae, meteors, and eclipses.

### 3.1 Cosmogony

In Palawa cultures, the sky, land, and people are intricately linked, and the stars form the basis of Palawa cosmogony (the formation of the world). The journals of Robinson indicate that Palawa spirituality is based on “star gods”, with a good spirit ruling over the day (Noiheener) and an evil spirit ruling over the night (Wrageowraper) (Ryan, 1996: 10). The creation of the world occurred when ancestral spirits formed the landscape, animals, vegetation, and sea. Traditions from across the island differ slightly, including the pronunciation of the names and details of the story (1979; Cotton 2013; McKay, 2001; Robinson and Plomley, 2008: 406-410), but in general they are similar.

Nuenonne man Leigh Maynard from Bruny Island details the story of the creation of Tasmania (Thompson and Tasmanian Aboriginal Community, 2011). It is unclear if this knowledge was passed to Maynard or if he is drawing from earlier written sources. Maynard describes the tradition as a circular story, like the cycles of the moon and the sun. Long ago, Tasmania (Trowenna) was a small sandbank in southern seas. Ice came and went and as the sea rose, the sun flashed fire. Punywin, the Sun man, and his wife Venna, the Moon, moved from horizon to horizon together, creating life and sinking into the seas each evening. But Venna could not travel as fast as Punywin, so he reflected light on her to encourage her move across the sky. During this time, the Sun and Moon were together in the sky. As Venna struggled to keep up with Punywin, he allowed her to rest on icebergs. One day the moon seemed to be permanently on the horizon. The day after, their first son, Moinee, was born. He was placed high in the sky above Trowenna as the Great South Star. The next day came their second son, gentle Droemerdeene. Punywin and Venna placed him in the sky between Moinee and themselves as the star Canopus. The day after Droemerdeene was born, the sun and moon rose together above the sandbank that was Tasmania. They dropped seeds for the trees and plants. The next day shellfish appeared in the waters and were plentiful. Troweena gradually rose from the seas and icebergs rubbed against Trowenna, pushing it from the great south land (mainland Australia) to the island we see today as Tasmania.

The Palawa guides on the Robinson expedition provide the first records of the creation traditions (Robinson and Plomley, 2008: 406). As in the Maynard account, the Sun and Moon are regarded as a man and woman, respectively. They give birth to two sons: Moinee (the elder) and Droemerdeenne (the younger). They came together to create the first man, named Palawa or Parlevar (now the name given to Aboriginal Tasmanians). Moinee first created Palawa with a tail like a kangaroo and no knee joints, making it impossible for him to sit or lay down. Seeing Palawa struggle, Droemerdeenne cut off his tail, then used animal fat to rub over the wound and gave him knee joints (*ibid*: 406). Droemerdeene and Moinee fight in the sky. Moinee was cast from the sky and lived on the Earth, followed by his wife, who went into the sea, and his children who came down as rain and fell into his wife's womb (ibid: 409). When Moinee died, he was turned into a stone found at Cox Bight. A Toogee elder, Timler, recounts a similar tradition (Cotton 1979).

Moinee is said to have made the first man, the rivers, and the islands - attributes also given to *Laller*, a small ant. The interchangeability of the two creator spirits may indicate they are one and the same: a totemic relationship similar to some practiced in mainland traditions (Robinson and Plomley, 2008: 406; Witzel, 2013: 11). The stellar identity of Moinee is not known. He is called the "Great South Star" who "comes out of the sea" (Robinson and Plomley, 2008: 406). Plomely identifies Droemerdeenne as Canopus, the second brightest star in the night sky. Canopus is circumpolar as seen from Tasmania and can appear to "come out of the sea" as it reaches its lowest altitude (~5.5° from southern Tasmania, which is close to the extinction angle) and begins climbing back up into the sky. Droemerdeenne was placed between Moinee and their parents, the Sun and Moon. This suggests that Droemerdeene is between Moinee and the ecliptic.

The identity of Moinee is unclear, as the star's Western counterpart is not named and the definition of "southern" is not explicit. Moinee is a bright star "in the south", presumably meaning southerly declination, and Droemerdeene is a bright star positioned between Moinee and the Sun and Moon, presumably the ecliptic. This leaves a number of options open. If we assume that the brothers are represented by the brightest stars in the night sky, then the best fit is Moinee as Canopus and Droemerdeene as Sirius. These two stars have similar right ascensions. If we connect a straight line between Canopus, Sirius, and the ecliptic, then Sirius lies almost halfway between Canopus and the ecliptic: $\Delta\alpha$ (Canopus-Sirius) = 36°, $\Delta\alpha$ (Sirius-Ecliptic) = 39°. Other combinations are possible, such as Achernar and Fomalhaut, but these stars are not as bright.

Another clue comes from Robinson and Plomley (2008: 425). On 1 August 1831, Robinson writes that Droemerdeene's brothers are two stars sitting south and east of Orion's belt:

> *"Tonight the Brune* [Bruny Island] *natives pointed out two stars to the southward, laying eastward of Orion's belt, which they said was Dromerdeenne and his brother, i.e. Beegerer and Pimerner. They were brilliant stars and appear to move towards the observer, rising as it were in the southern horizon and setting in the north."*

Plomley identifies these two stars as Betelgeuse and Sirius. He suggests the text may be in error and should read "Dromerdeene's *brothers*, i.e. Beegerer and Pimerner", instead of "Dromerdeenne and his brother" (Robinson and Plomley, 2008: 500). The recorded traditions do not mention additional brothers of Moinee and Dromerdeenne. Robinson's journal indicates a single brother with two variations in name: Beegerer and Pimerner. The passage is confusing. Did he mean "*Beegerer or Pimerner"?* Neglecting small long term changes due to proper motion, the declination of stars is constant meaning a star rising in the southeast will set in the southwest, never the northwest. What does he mean? Are the names Beegerer and Pimerner some variation of Moinee?

If we assume Robinson is recording different names of a single brother of Dromerdeenne, (whom we identify as Moinee) then his description of the two stars "laying eastward of Orion's Belt" and rising in southward (southeast), best fits with Canopus and Sirius. Orion was not visible until the early morning on the day Robinson wrote in his journal (1 August). When it did rise, Sirius and Canopus are clearly visible in the southeastern sky (the former rising at nearly the same time as Orion's Belt and the latter already 16° above the horizon). Both stars move in a northerly direction until the sun rises and the stars disappear. By this time, Sirius is in the northeastern sky while Canopus remains in the southeast.

We suggest the evidence best supports the identities of the star-brothers as Canopus (Moinee) and Sirius (Dromerdeenne). We feel it is a better fit than Dromerdeenne as Canopus and Moinee as an unidentified star, but this remains uncertain.

On a final note, the idea that a bright star appeared in the sky but is no longer visible hints to a possible supernova event, but there is currently no supporting evidence for this interpretation (Hamacher, 2014).

### 3.2 The Gemini Twins and the Origin of Fire

Traditions that describe how fire was brought to the Palawa tend to focus on the actions of two ancestor spirits who can be seen today as two stars near the Milky Way. A tradition from Oyster Bay tells how the two men stood on a mountaintop and "threw fire, like a star" that "fell among the blackmen" (Milligan 1859: 274). The two men live in the clouds and can be seen in the night sky as the stars Castor and Pollux (the Gemini twins in Greek traditions). On 14 August 1831, Robinson discussed religion with Mannalargenna. Mannalargenna said that two men created fire and now live in the skyworld. Mars was his [*sic*] foot and the Milky Way his [*sic*] road. According to Mannalargenna, the Cape Portland people believe fire was first made by Pormpenner. This name will be mentioned twice more in relation to fire but will be spelt differently each time as it was recorded: *Pardedar* (Robinson and Plomley, 2008: 872) and *Parpedder* (*ibid*: 577). On 15 August 1831, Robinson wrote that Palawa from Bruny Island said two stars in the Milky Way represent two men (*ibid*: 433). Woorrady attributes *Parpedder* as being the one who gave fire to the people of Bruny Island. Why two men are identified, but then refereed to in the singular is unclear.

On 16 August 1831, Mannalargenna called the two stars Pumpermehowlle and Pineterrinner. He describes them as the two spiritual ancestors who created people and fire, but the stars' Western counterparts were not named. Later, Milligan (1859: 274) recorded a tradition from an unknown Oyster Bay person, who identified Castor and Pollux (the Gemini twins) as the two men who create fire. In Boorong traditions of western Victoria (Stanbridge, 1858: 140), Castor (*Yuree*) and Pollux (*Wanjel*) represent two young male hunters that pursue a kangaroo and kill him at the commencement of the "great heat" (summer).

There is a problem with setting a planet as the body-part of a celestial ancestor. Planets constantly move. Was the foot of the man (men?) actually Mars, or a red star of similar brightness? During 14-16 August 1831, Castor and Pollux heliacally rose. They set before dusk, so are not visible in the evening sky. Mars is in near conjunction with Saturn (and $< 2°$ distant) at very low altitudes at dawn, with Venus and Mercury above them in the western sky. Since the stars were not in the sky when Robinson was told about them, how does he identify the foot of the man as Mars?

Castor and Pollux are northerly stars, only reaching a maximum altitude of ~16° and ~20°, respectively, as seen from Tasmania. There are no bright (first magnitude) red stars between the Gemini twins and the Milky Way. Orion is on the other side of the Milky Way and the ecliptic passes between them. Mars could appear at the "foot" of the hunters walking on the Milky Way, but this will be (relatively) sporadic. Earlier in May 1831, Mars was visible between the Gemini twins and the Milky Way. Perhaps this is the reason Mars is recorded in this way? The stellar counterparts remain unclear, but this is the only written record of the hunters' identity.

If the recorded information is from from Mannalargenna, then from who did Milligan get his information? Milligan was a doctor on Flinders Island after the Robinson expedition and would have formed relationships with the same Aboriginal people who accompanied Robinson. Mannalargenna died in 1835, nine years before. Sometime between 1843 and 1855, Milligan recorded *Legend of the Origin of Fire.* Milligan does not specify the gender of the narrator. Still, a census was performed by Robinson in 1836 renaming Aboriginal people with English names (Plomley and Robinson, 1987: 878). It is probable this list contains the name of the person who told this tradition. It is important to note among this role-call are Wapperty, Calamarowenye, Truganini, and Bullrer - all of whom were on the Robinson expedition and originate from the Oyster Bay region (Gough, 2014: 33). Any of the aforementioned people could have told this story to Milligan.

### 3.3 The Coalsack Nebula and the Celestial Stingray

Robinson records a tradition of a stingray in the sky on 13 March 1834 at 23:00 (Robinson and Plomley, 2008: 895). The stingray is seen as a black spot in the Milky Way (or Orion's Belt) that people are spearing. It is called *Larder* in the south and *Larner* on the east coast. Robinson uses *Larder* in 1831 to identify the "dark area" in the Milky Way (ibid :497). *Larner* is also used in relation to Mars and *Lawway Larner* translates to "Milky Way/road - Stingaree" (*ibid*: 895). *Larner* may have been incorrectly linked with Mars, or this word may take on other meanings. Another version of this word for fish is '*Lerunna*', recorded by Milligan (1890: 28) as "Flat Fish or Flounder."

Robinson identifies the "black spot" as being in the Milky Way or Orion's Belt. This is probably the Coalsack (Figure 4), a dark absorption nebula that can be seen clearly with the naked eye and appears as a dark hole against the backdrop of the otherwise bright Milky Way.

The Coalsack borders the Western constellations of Centaurus, Musca, and Crux, not Orion (which is 90° on the other side of the sky). On 13 March 1834, Orion is sitting prominently above the western horizon at 23:00 and does not contain any large or obvious dark nebula visible to the naked eye.

Robinson's entry states that the stingray is being speared by the men. We suggest the spears may be the Pointer Stars, Alpha and Beta Centauri. In northwestern Victoria, the Coalsack is an emu named *Tchingal* in the Wergaia language. The eastern stars of the Southern Cross (Alpha and Beta Crucis) are the pointy ends of the spears of two warriors who speared the emu through the neck and rump (Stanbridge, 1858: 139).

*Table 2: Notes regarding star names found at the end of Robinson's journals, April–July 1831.*

| Object | Oyster Bay | Brune/ Bruny Island | Cape Portland |
|---|---|---|---|
| Mars | | LAW.WAY LAR.NER | LAW.WAY DEVER.ER |
| Star (1) | PUCK.AR.NE.PEN.NER | PY.LE.BAY | PUM.PER.ME.HOWL.LE |
| Star (2) | LORE.NE.PEN.NER (wife) | LAW.WAY | PINE.TER.RIN.ER |
| Black Milky Way | | LAR.DER | PY.ER.DREEM.ME TONE.NER.MUCK.KEL.LEN.NER |
| White Milky Way | | LAW.WAY.TEEN.NE | PUL.LEN.NER |

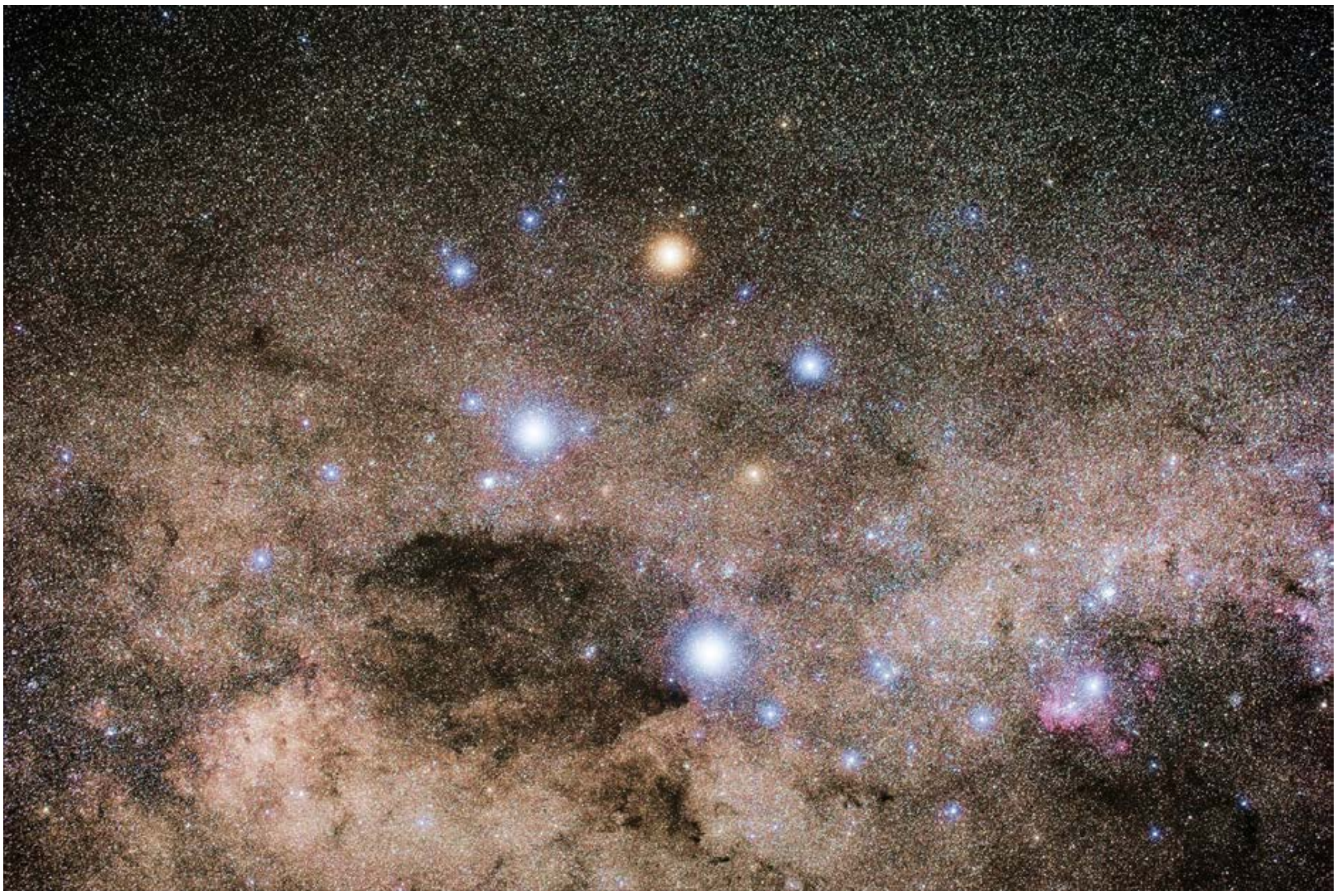

*Figure 4: The Coalsack, bordering the Southern Cross (Crux). Image: Wikipedia Commons License.*

Cotton (2013) provides a retelling of the story of the *Legend of the Origin of Fire*, in which the men and their wives are identified as the stars of Crux (the Southern Cross). The original account of the story was recorded by Joseph and Isobel Cotton from Timler, an East Coast Palawa storyteller, in the 1830s but the records were lost in a house fire in 1959 (Stephens 2013). Years later, descendent William Jackson Cotton (1909-1981) rewrote the stories from memory and published them as *Touch the Morning* (Cotton 1979). More recently, William Cotton's daughter, Jane Cooper, published William's recollected stories in Cotton (2013), but without consultation with the local Aboriginal community (Johnson et al, 2015:14). The re-published version of the story, called *Cross of Fire* evokes substantial Christian imagery but identifies the mountain in the story, Meledna Lopatin (Mountain of Fire), as Mount Amos in eastern Tasmania (Cotton 2013: 62). It also names the two men who bring fire as *Una* and *Bura*.

Translations of the mens' names are found in the vocabularies of Roth's (1890) consolidated vocabulary. *Una* or *Une* translates to *fire*. The two words joined *Une Bura* is translated to *lightning* and *Bura* alone is translated as *thunder* (*ibid*: xiv). *Una* and *Bura*, different to the names given to these two stars in Robinson's recordings, give a literal and direct translation to their intended meaning: *fire*, *lightning*, and *thunder*. Similar to *Legend of the Origin of Fire*, this story ends with the two men and the two women returning to the sky. However, *Cross of Fire* identifies the four stars as the the brightest four stars in Crux (the Southern Cross), called or *Urapane Lopatin* (Cross of Fire) in the eastern Palawa language (Roth 1890: xxiii, xi, respectively). In this account, the stingray joins them in the sky. This is the first account that mentions this. Conversely, the *Cross of Fire* story does not identify the Coalsack or any dark patches in the sky (see Cotton 2013: 71). It is difficult to ascertain the accuracy of these records, given that they were retold from memory long after they were recorded and contain heavy Christian imagery.

Going back to Robinson's journals between April and July 1831, he identifies a particular star, *Lorenepenner,* as being female (Table 1). The Oyster Bay word *Lorenepenner* is translated as *wife* (Robinson and Plomley, 2008: 497). Milligan (1890: 51) identifies the women as *Lowanna*; a common word for women in Milligan's own collected vocabulary. The presence of a female-star in Robinson's notes supports the idea that a version of *Legend of the Origin of Fire* could have been relayed to Robinson during his mission, with the names of the stars representing the names of the ancestral protagonists featured in this tradition.

On 27 June 1831, Robinson (Robinson and Plomley, 2008: 497) writes:

> *"In conversation with the natives respecting the stars. These people, like the ancients, have described constellations in the heavens as resembling men and women, men fighting, animals, and limbs of men; together with names for the stars. The Aborigines pointed them out."*

Unfortunately, like many of Robinson's entries, it is condensed and short on detail. The excerpt provides a summary of features of Aboriginal interpretations of the stars, many identifiable within Milligan (1890).

This tradition, shared by a member of the Oyster Bay group, can be unpacked beyond that of labels and language. Oral traditions are passed on for (potentially) thousands of years. These traditions are encoded with information significant to the survival and navigation of the physical and social landscape. Reading the canopy of stars above as a form of traditional text informs practice on land, which is evident in *Legend of the Origin of Fire.* On the surface this tradition explains how fire came to the people of Tasmania. This story contains information about seasonal indicators, fishing customs, burial and healing practices, as well as fire attainment.

Palawa women living on coastal environments, like Oyster Bay, spent many hours in the water. Acting as the prime hunters of shellfish and being taught from a young age, they could dive considerable depths on a single breath (Robinson and Plomley 2008: 66-88; Johnson et al, 2015: 39). Due to the significant time people spent in the sea, the oral tradition and the night sky were used to inform cultural practices regarding how to navigate the oceanic environment safely.

According to the oral tradition, two women diving for crayfish were 'sulky' due to their unfaithful husbands. Consequently, the women were speared by the stingray and died. The same wording was used in an earlier recording of a separate incident in Robinson's journals. On 4 November 1830, Robinson describes the women returning from diving for crayfish off Swan Island, where they were chased by a shark (Robinson and Plomley, 2008: 302). The women were described as sulky, which made the sharks come.

*Legend of the Origin of Fire* also describes the women as being "sulky" when they were speared and killed by the stingray. Afterward, the two star men arrive and kill the stingray with their spears (Milligan, 1890: 13). This reflects culture practiced on the ground. Lloyd (1862: 52) recorded a personal observation from an Aboriginal hunting trip the morning after a significant corroboree was held during a full moon. Up to 300 people surrounded stingrays in a semi-circle and men speared them.

The final section of this tradition (Milligan, 1890: 13) explains the revival of the two women who were speared by the stingray. The dead women were placed on either side of a fire and the men placed "blue ants" on the breasts of the women. After being bitten severely, the women came back to life. The importance of fire within Aboriginal culture and its relationship with rejuvenation and healing is described in tradition. On the Wellesley Islands in the Gulf of Carpentaria, meteors signal the end of a healing process (Cawte 1974: 110). A disease called *malgri* is treated by lighting a fire next to the patient, encouraging them to sweat. Similarly, Bonwick (1870) describes a Palawa patient drinking lots of cold water then lying by the fire to encourage perspiration.

The blue ant (*Diamma bicolor*) is actually a parasitic wasp found throughout southeast Australia. The female has an ant-like appearance and if disturbed, her stinger can cause burning pain and swelling. Early recordings ascribe large ants or *Diamma bicolor's* eggs as being a delicacy among Palawa (Noetling, 1910: 281). Blue ants are active in mid to late summer, playing an important role in pollinating native plants, a possible timing component indicating seasonal change within the tradition (ibid).

### 3.4 Time, Navigation, and Astronomy

Time-keeping was important for food economics, calendar development, and ceremony. Consolidated vocabularies of the language groups of Tasmania (Table 2) reflect words used to indicate time of the day (e.g. sunrise, midday, sunset, twilight), astronomical presence (e.g. starlight, moonlight), and seasons (Milligan, 1890; Plomley, 1976). There are no words for the concept of time itself, an observation made by Stanner (2011).

Roth (1890: 146) makes a fleeting comment on the understanding of time and astronomy of Aboriginal people. He notes that they point out the diurnal motion of the sun with their hands and hold up two fingers to denote two days. He then claims that "*This is the only reference to any knowledge of the movement of the heavenly bodies.*" Conversely, Robinson writes on 13

March 1834 that the Palawa are quite familiar with the stars and have names for them all and are aware of their movements (Robinson and Plomley, 2008: 302). This record demonstrates "knowledge of the movement of heavenly bodies."

On 25 December 1830, Robinson praises the "considerable knowledge" of Palawa on meteorology so much, that they had "attained to such celebrity." As a result, Robinson and "white men" in general would consult them on the subject and be pleased at the information as it "seldom found them to err" (*ibid*: 334). Palawa used the stars and clouds to determine when to fish, build huts, and travel.

*Table 2: Vocabulary table of Aboriginal words indicating time of day, from Plomley (1976).*

| Topic | Bruny Island /Southern TAS | Oyster Bay | Northern TAS | Western TAS |
|---|---|---|---|---|
| **Twilight** | nunto neenah | teggrymony keetana narra long - boorack | | |
| **Early morning twilight** | nunawenapoyla | tuggamarannye | | |
| **Sunrise** | panubre roeelapoerack | muenattemelar | warkala wetinneger | |
| **Sunrise** | | puggalena parrack boorack | | |
| **Midday** | toina wunna | tooggy malangta | | |
| **Midday** | wer | | | |
| **Sunset** | punubra tongoieerah | wietytongmena | | |
| **Sunset** | | partopelar | | |
| **Moonlight** | weetapoona | wiggetapoona | | weenapooleah |
| **Starlight** | oarattih | teahbertyacrackna | | |

Astronomical knowledge is embedded through Aboriginal traditions, but not always as obvious as a direct comment. Robinson's manuscripts (Plomley, 1976: 51) describe a song from the northwest, north coast, and interior Palawa groups, that is possibly used for navigation and travel. The Palawa "repeat the words *tonener* (sun) and point the way the sun is travelling in her course, and point to where they are stopping for the sun to be there." *Tonner* is also refers to "West" (*ibid*: 205) and is part of the word for the Black Milky Way; *tonnermuckkellenner* (*ibid*: 408). The description of the actions that accompany the repetition of *tonner,* indicates this song was sung to help with timing and navigation on their journey, serving as an insight into a Tasmanian songline. We suggest that the songline describing "the way the sun is travelling" indicating "where they are stopping" in relation to the sun, demonstrates a form of celestial navigation.

### 3.5 Seasonal Change and Astronomy

Three unidentified stars relating to seasonal change are mentioned three times in the literature. Twice in Robinson's journals: on 20 June 1832 and 30 June 1834 and again in interviews conducted by Ernest Westlake between 1908 and 1910. In all instances, the three stars are used to track time seasonally. In the 1832 account, the dark phases of the moon are used in conjunction with stars to indicate specific shorter intervals of time. The identities of these stars remain unknown.

Robinson's 30 June 1834 entry (Robinson and Plomley, 2008: 111) provides the positions and magnitudes of the mystery stars and their use as a seasonal indicator (names listed in Table 3):

> *"AM, calm and clear, fine weather, sun hot. The natives showed me the three stars which they say is a sign that the fine weather is coming and when those stars are vertical the fine weather is come. They appeared in the heavens to the eastward. No. 1 was large and is called the mother, No. 2 the husband is of lesser magnitude and No. 3 the offspring is hardly visible. They are called by the Brune natives PUR, by the western natives LONE.ER.TEN, by the northern natives NOE.GO, and by the natives of Oyster Bay PARNG.ER.LIN.NER."*

*Table 3: Names given to the three stars shown to Robinson on 30 June 1834.*

| Origin | Aboriginal Word | Possible meaning |
|---|---|---|
| Bruny Island | PUR | White Edible Berry |
| Oyster Bay | PARNG.GER.LIN.NER | Wife (Eastern) |
| Northern | NOE.GO | West Point (place) |
| Western | LONE.ER.TEN | Wife |

The Bruny Island word *Pur* is similar to *Purrar*, a Bruny word given to white edible berries (Plomley, 1976: 340). This association suggests that the star would appear white, ruling out red stars. The Western group's word *Loneerten* has connections with *Looner*, or "wife" across many Palawa language groups (*ibid*: 471). The Northern word *Noego* is quite close to *Nongor*, the Palawa name for West Point in northern Tasmania. *Parnggerlinner*, from Oyster Bay, may be related to the word *Parnuneninger* for "wife" given by some eastern groups (*ibid*: 321).

Key information provided from this journal which aide in identifying the stars is as follows:

(1) The date visible was 30 June 1834 at dawn (see Figure 5),
(2) The stars appeared eastward (azimuth between 0° and 180°),
(3) The stars are of different magnitudes: a large (bright) star (presumably first magni-

tude), a lesser bright star (presumably a second or third magnitude star), and a hardly visible star (presumably fifth or fainter magnitude).

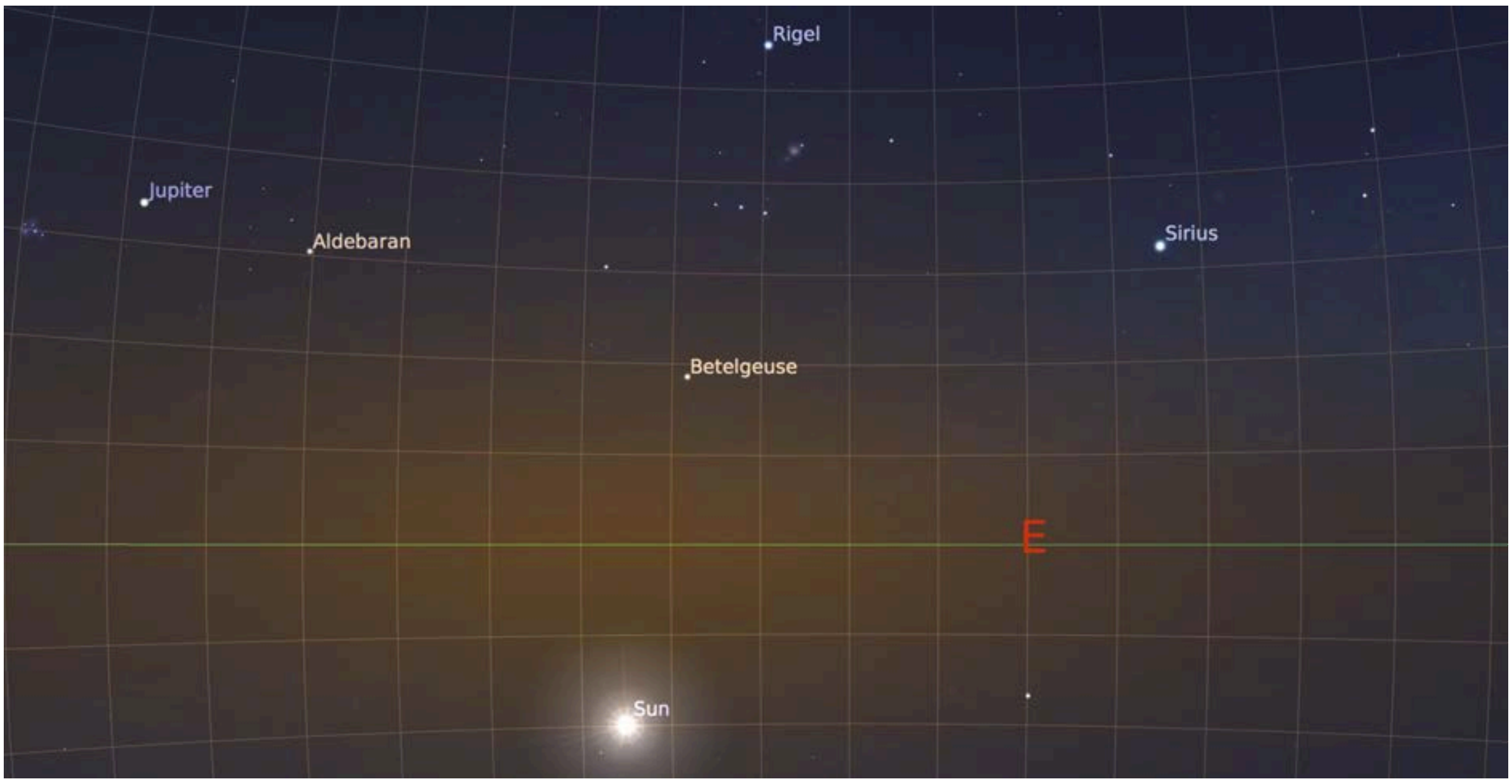

*Figure 5: Stars in the eastern sky at dawn (when the sun is at an altitude of -10°) on 30 June 1834. Notable first magnitude stars are Sirius (right), Rigel (top), Betelgeuse (centre), and Aldebaran (left). The grid is shown in 5° increments in both declination and right ascension. Image: Stellarium.*

The orientation of the three stars in his drawing is an illegible number. Plomley interpreted this number to be 30, presumably from looking at the sketch drawn by Robinson in Figure 6. Robinson writes that the stars indicate fine weather is coming. When they are vertical, fine weather has come. Identifying a period of "fine weather" in the calendar year will approximate a date to then test for stars that are vertical at this time. The clan territories that name the stars are from the North, East, West, and South of the island, indicating that 'fine weather' would (on average) be experienced across the whole of Tasmania. Meteorological data shows that Tasmania experiences constant rainfall through the year, with winter having the most, and can experience multiple seasons in one day. Based on this data, "fine weather" could be considered the "summer" months, most likely January. Since climate records have been kept, January sees the least amount of rainfall [1].

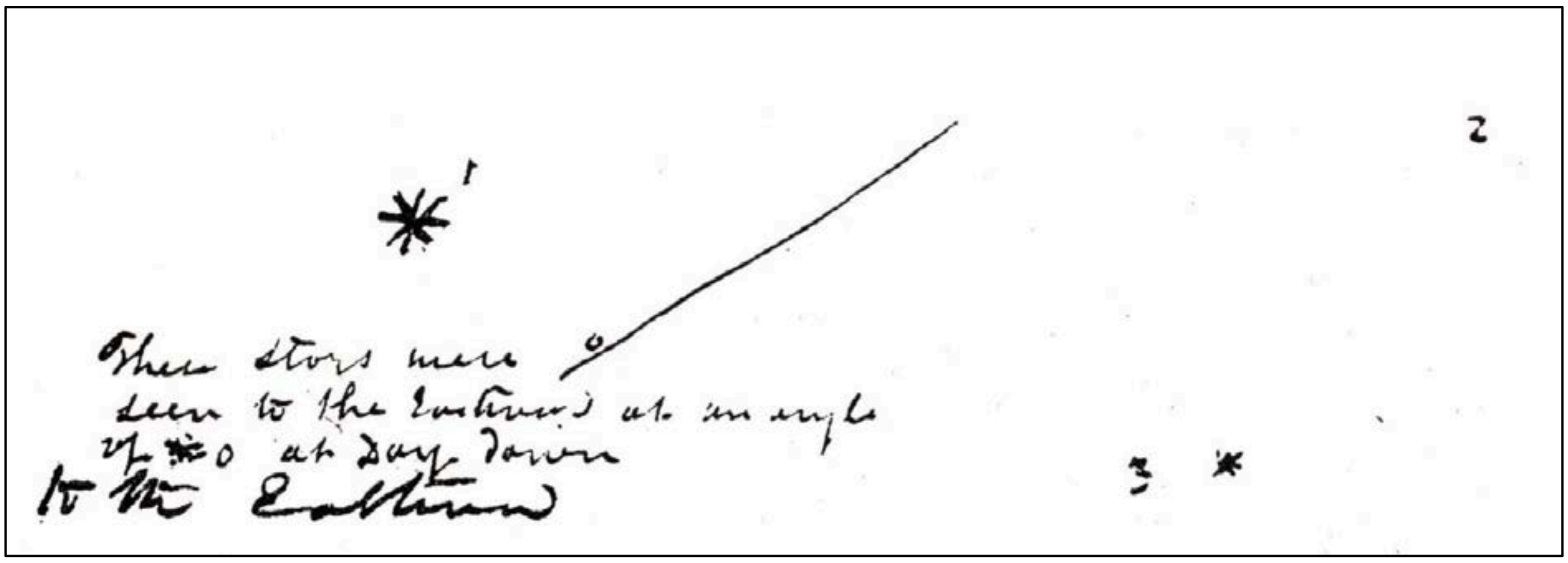

*Figure 6: The sketch Robinson made showing the orientation of the three stars described on 30 June 1834.*

With these variables in mind, Plomley's identification can be tested. Plomley classified the three stars as the Pointer stars, Alpha and Beta Centauri (Plomley, 1997; Robinson and Plomley, 2008: 953). On the morning of 30 June 1834 (when the sun is 10° below the horizon: see Hamacher, 2015), the Pointer stars appear in the southwest (Az = 185°-190°), not the east. The Pointer stars are both of similar brightness and a third barely visible star in a line with these is difficult to identify. The Pointers are circumpolar as seen from Tasmania, so they can appear to be horizontal on some occasions in the morning and vertical in others.

The Pointer stars are also of similar brightness ($V_{mag}$ = +1.33 and +0.61 for Alpha and Beta, respectively). The Pointers are vertical (having the same azimuth) in the morning sky in mid-January and are horizontal (having the same altitude) high in the sky a month later in mid-February, or horizontal low in the sky in late July/early August.

Plomley's identification is inconsistent with the information drawn from the journal entry. Despite lining up vertically at certain times of the year, Alpha and Beta Centauri encircle the south celestial pole. Robinson states that the three stars sit eastward. In subsequent publications, researchers have mis-transcribed Plomley's hypothesis by claiming the stars in question are Alpha and Beta Crucis (two brightest stars in the Southern Cross), causing confusion (Coon, 1972: 288).

While it is difficult to accurately label the stars from the description given, there are a number of stars sitting on the eastern horizon on that morning. Many move to a vertical position on the western horizon, on a mid-January morning. The following stars/star groups appear prominently in the east at dawn on 30 June 1834:

- The Pleiades star cluster (Messier 45)
- Sirius (Canis Major)
- Aldebaran (Alpha Taurii)
- Betelgeuse (Alpha Orionis)
- Bellatrix (Gamma Orionis)
- Orion's Belt (Mintaka, Alnilam, Alnitak).

This list identifies the most prominent stars visible at this time. Canopus is not included in this list as it sits closer to southeast, never sets below the horizon (circumpolar), and has been previously identified as creator ancestor Droemerdeene (although we question this identification). We attempt to identify the stars recorded by Robinson by examining a suitable list of candidates and comparing them with the information provided in the record, utilising the stars' magnitudes, relative positions, and colours (Tables 4 and 5).

Robinson's sketch does not show the stars in a linear pattern. The third star is described as being "barely visible". Due to the faint magnitude of the third star there are multiple candidates. When grouping the three stars we take into consideration the orientation of the third star as well as the magnitude; only picking stars that were fifth magnitude or brighter.

*Table 4: Seven possible groupings of three stars as recorded by Robinson. The groupings of three stars are given by common name, Bayer designation, visual magnitude (Vmag), general spectral type (colour), and the star's coordinates (right ascension and declination in J2000).*

| **Common Name** | **Bayer Designation** | **Vmag** | **ST (Colour)** | **RA (J2000)** | **DEC (J2000)** |
|---|---|---|---|---|---|
| Rigil Kent | α Centauri | 0.01 | G (Yellow/white) | $14^h\ 39^m\ 36.5^s$ | –60° 50′ 02.4″ |
| Hadar | β Centauri | 0.61 | B (Blue) | $14^h\ 03^m\ 49.4^s$ | –60° 22′ 22.9″ |
| Hip 70264 A | n/a | 4.90 | K (Orange) | $14^h\ 22^m\ 38.02^s$ | –58° 27′ 36.7″ |
| Mintaka | δ Orionis | 2.23 | O/B (Blue) | $05^h\ 32^m\ 00.4^s$ | −00° 17′ 56.7″ |
| Alnitak | ζ Orionis | 1.77 | O/B (Blue) | $05^h\ 40^m\ 45.5^s$ | −01° 56′ 33.3″ |
| Alnilam | ε Orionis | 1.69 | B (Blue) | $05^h\ 36^m\ 12.8^s$ | −01° 12′ 06.9″ |
| Sirius | α Canis Majoris | −1.46 | A (Blue/White) | $06^h\ 45^m\ 08.9^s$ | −16° 42′ 58.0″ |
| Adhara | ε Canis Majoris | 1.50 | B (Blue) | $06^h\ 58^m\ 37.6^s$ | –28° 58′ 19.0″ |
| Wezen | δ Canis Majoris | 1.82 | F (White) | $07^h\ 08^m\ 23.5^s$ | −26° 23′ 35.5″ |
| Aldebaran | α Tauri | 0.87 | K (Orange) | $04^h\ 35^m\ 55.2^s$ | +16° 30′ 33.5″ |
| Bellatrix | γ Orionis | 1.64 | B (Blue) | $05^h\ 25^m\ 07.9^s$ | +06° 20′ 58.9″ |
| Meissa | λ Orionis | 3.50 | B (Blue) | $05^h\ 35^m\ 08.29^s$ | +09° 56′ 03.0‴ |
| Betelgeuse | α Orionis | 0.42 | M (Red) | $05^h\ 55^m\ 10.3^s$ | +07° 24′ 25.4″ |
| Mirzane | β Canis Majoris | 1.99 | B (Blue) | $06^h\ 22^m\ 42.0^s$ | −17° 57′ 21.3‴ |
| Beta Monocerotis | β Mon | 4.60 | B (Blue) | $06^h\ 28^m\ 49.16^s$ | −7° 01′ 58.2‴ |
| Rigel | β Orionis | 0.12 | B (Blue) | $05^h\ 14^m\ 32.26^s$ | −8° 12′ 06.0″ |
| Mirzam | β Canis Majoris | 1.99 | B (Blue) | $06^h\ 22^m\ 42.0^s$ | −17° 57′ 21.3″ |
| Saiph | κ Orionis | 2.09 | B (Blue) | $05^h\ 47^m\ 45.4^s$ | −09° 40′ 10.6″ |
| Bellatrix | γ Orionis | 1.64 | B (Blue) | $05^h\ 25^m\ 07.9^s$ | +06° 20′ 58.9″ |
| Saiph | κ Orionis | 2.09 | B (Blue) | $05^h\ 47^m\ 45.4^s$ | −09° 40′ 10.6″ |
| Gamma Monocerotis | γ Mon | 3.95 | A (White) | $06^h\ 14^m\ 51.10^s$ | −06° 16′ 26.0″ |

*Table 5: Possible groupings of three stars using the description recorded by Robinson. The groupings are tested to see if they match the criteria using Robinson's recorded descriptions.*

| Star Groupings | Non-Red Stars | Eastward | Dawn | Sketch and magnitude | Vertical in mid-January |
|---|---|---|---|---|---|
| Group 1 | ✔ | | ✔ | | ✔ |
| Group 2 | ✔ | ✔ | ✔ | | ✔ |
| Group 3 | ✔ | ✔ | ✔ | ✔ | ✔ |
| Group 4 | | ✔ | ✔ | | ✔ |
| Group 5 | | ✔ | ✔ | ✔ | ✔ |
| Group 6 | ✔ | ✔ | ✔ | | ✔ |
| Group 7 | ✔ | ✔ | ✔ | ✔ | ✔ |

Groups 3 and 7 are the only ones to meet all of the criteria. The Group 3 stars (Figure 7) fit the description reasonably well. Of the three stars, Sirius the brightest star in the sky, sits eastward at dawn. Its orientation with Adhara and Wezen are similar the sketch drawn by Robinson (Figure 6). Sirius, the most northerly of these stars, exceeds an altitude of 5° (the extinction angle) when the sun is at −10° altitude on 15 June 1832. This is considered the first day the three stars of Group 3 are unambiguously visible in the east at dawn. The relative brightnesses are roughly consistent, although Wezen, with a Vmag of 1.2, is not "hardly visible". But when the stellar trio are very low on the horizon at dawn, the background light is enough to sufficiently obscure it and make it appear much fainter.

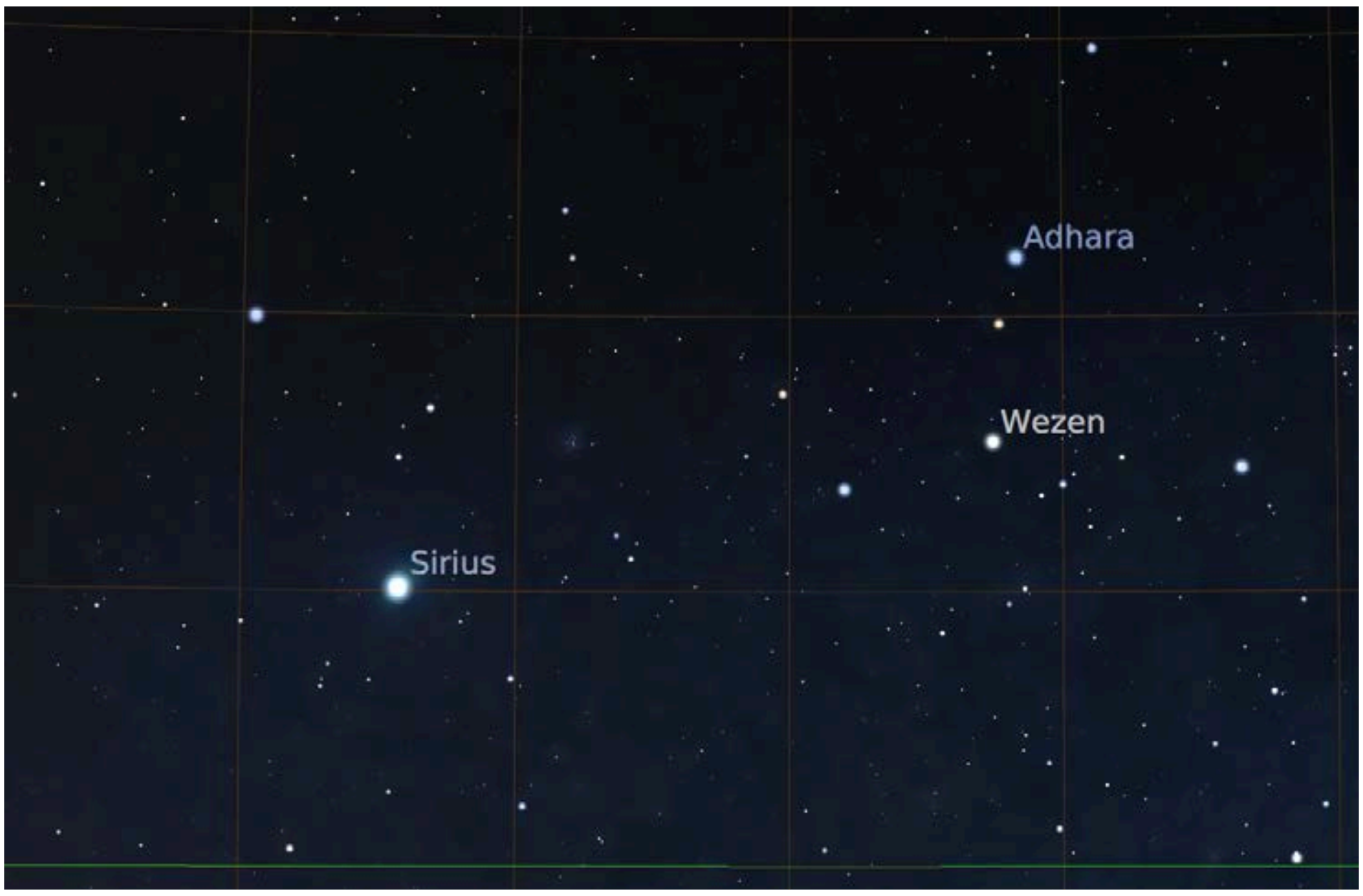

*Figure 7: A trio of bright stars in Canis Major - Sirius, Adhara, and Wezen (Group 3) - are the best fit for the three stars described and illustrated in Robinson's journals. Sirius has an azimuth of 107.5° at an altitude of 5° at dawn (when all three stars are first visible together above the horizon) on 15 June 1832 (shown). Image: Stellarium.*

Group 7, while meeting all of the criteria, seems less likely as they straddle Orion's Belt. Orion's Belt is mentioned by Robinson in earlier entries, indicating he knew the asterism and is likely to have labeled or located them when describing the three stars. There is one problem: if Plomley is correct in identifying Betelgeuse and Sirius as Dromerdeene's brothers, would he not recognise Sirius - the brightest star in the sky - and realise it is already identified in Palawa traditions?

An altercation between the Tarkiner group of Northwest Tasmania and the Robinson party occurred and a fight was scheduled to take place at Nongor (West Point). The below two entries, on 19 and 20 June 1832, were made by Robinson regarding the timing of this fight (Robinson and Plomley, 2008: 652):

> **19 June 1832**: "*I learnt that the TARKINER natives were to come and fight them when the rest came back from Robbins Island - the TARKINER would come two dark nights after the moon was gone (it was now moonlight)."*
>
> **20 June 1832**: "*Learnt that the greater part of the natives had gone to Robbins Island and were engaged in getting spears, that they would return again when two darks or when the three stars come."*

The new moon (denoting the days where less than 3% of the moon facing the earth is illuminated) occurred from 26-28 June 1832. The fight with the Tarkiner is scheduled on 29 June 1832 - two nights after the "disappearance" of the moon. This allowed themselves nine days to prepare. Additionally, the appearance of the stars may signify the time of day, not the day of the month. This may have been indicated by "two dark nights" (date) and "when the three stars come" (time), translating to Friday, 29 June 1832 at approximately 05:45.

"When the three stars come" seems to describe the three stars we are trying to identify in this section. Coincidently, Robinson was shown the three stars that indicate seasonal change exactly two years later on 30 June 1834, suggesting the same three stars are being described in both accounts. The earlier mention of the three stars in 1832 indicates that they have not yet appeared in the sky.

In an interview with Ernest Westlake, noted under the heading 'Springtime', Augustus Smith (Fanny Cochrane Smith's grandson) spoke of three stars (Plomley et al., 1991: 63):

> *"Three little stars in the east on a level only once in a year. Thought a lot of them, just to see them blinking. FS thought it a terrible thing if didn't welcome these three little stars. Would sprinkle the ashes from the hearth very early in the morning before the sun had risen, when the stars are bright."*

Like the two previous entries in Robinson's journals, Smith describes the three stars to the east in the early morning 'on a level', and associates them with seasonal change (springtime). The three stars of Orion's Belt (Alnitak, Alnilam, and Mintaka) are seen in the early morning sky rising 'on a level' during the winter months of June, July, and August (Figure 8). The stars of Orion's Belt rise heliacally (at dawn) around 8 June each year, and heliacally set (at dusk) around 19 July. The heliacal rise appears to be premature for a welcoming of Spring, yet the Orion asterism is visible in the sky, just as described in the three literature entries.

The earliest recording of three stars in June 1832 gives timing components (dates, two dark nights, indication of moonlight) to cross check. The emergence of the three stars of Orion's belt is in line with the description. Orion's belt appears above the horizon early in the morning from 05:30, before setting with the sun at 07:00.

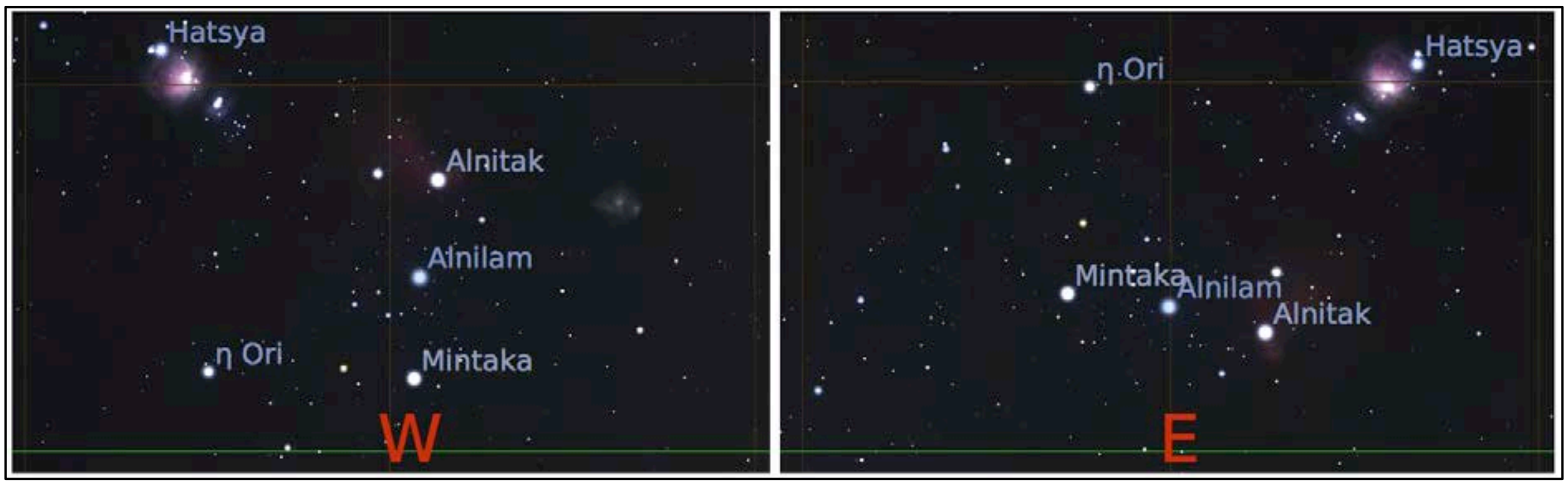

*Figure 8: The stars of Orion's Belt rising in the East (right) and setting in the West (left) as seen from Tasmania in the 1830s. The belt stars are "level" as the rise, and perpendicular to the horizon as they set. Image: Stellarium (set "without atmosphere" for clearer visibility for the reader).*

Two years later, nearly to the day, Robinson writes about three stars again. Accompanying the entry is the sketch placing the stars as 'eastward' at 'day dawn'. The three stars of Orion's Belt are clearly visible on the horizon in the east at this time. The repetition of the position of the three stars at the similar time of year supports the idea that the three stars of Orion's Belt could be the stars recorded in the journal. The constellation of Orion is visible above the horizon during summer nights, supporting the idea that the first appearance of these stars would be welcomed after a cold Tasmanian winter. This is uncertain, but is the topic of future ethnographic work.

### 3.6 Lunar Traditions

As noted in Cotton's (1979; 2013) recording of Toogee elder Timler's tradition, the Sun and Moon are parents of the creation ancestors that became the first stars. Similar traditions exist across Tasmania. The Palawa of Bruny Island told Robinson how the moon-woman, Vetea, got her dark patches (mare on the Moon; Robinson, 1831: 412):

> *"The Brune natives affirm that the moon (VETEA) came from England and that's she stopped at the RORE.DAIR.RE.ME.LOW, that is, the country at Oyster Bay, that the kangaroo and mutton fish asked the moon to stop there, that the moon was LOONER, woman, that she was roasting mutton-fish when the sun (PARNUEN) came and swept her away, and she tumbling into the fire was hurt on her side and then rolled into the sea, and afterwards went up into the sky (WARRANGERLY) and stopped there with her husband the sun. They say the rainbow is the sun's children.* [Woorady] *Told me if I looked I would see it black where she had been burnt."*

The Adnyamatana of the Flinders Ranges in South Australia have a tradition in a similar vein. Vira, the moon-man, falls off his stick ladder while trying to punish his nephew for stealing his food (Tunbridge, 1988: 68-69). On impact he burst open, leaving marks on his belly.

The Moon was used by Palawa to tell the time (Robinson and Plomley, 2008: 652) and count

(*ibid*: 267). The appearance of the Moon could also signal a change in weather (*ibid*: 334):

> *"...if a circle* [halo] *is round the moon it's a sure sign of bad weather. Indeed they have numerous signs by which they judge and I have seldom found them to err. Thus they are enabled to know when to build their huts, to go to the coast for fish, travel etc. They also judge by the stars and have names by which they distinguish them."*

In the weather folklore of cultures around the world, lunar haloes have long been used to predict bad weather (e.g. Guiley 1991: 22). The halo itself is caused by moonlight being refracted by ice crystals in the atmosphere. These form in cirrus clouds, which often come before a low pressure system, of which rain is a frequent result.

These traditions emulate a constant theme of disruption and restoration that is common in lunar traditions. We argue that the moon as a symbolic cycle of pain and healing that is reflected on the bodies of Palawa. Scarring was first thought to be unique to each group, as a distinguishing feature between nations. Yet often when there is mention of cicatrices, Robinson offers an astronomical motif in partnership, indicating meaning beyond the cosmetic (Johnson et al., 2015: 35). Sightings of moon or crescent shaped markings on bodies appear, but are not limited to the east coast of Tasmania. Lieutenant Le Paz, a member of French explorer Marc-Joseph Marion Dufresne's expedition in 1772, noticed "several little scars or black marks in a crescent shape" on the chest of a young man when they landed on the east coast of Tasmania (Duyker, 1955: 33). On 1 November 1830, Robinson observes most of the people from the eastern groups "had the form of the moon cut on their flesh" (Robinson and Plomley, 2008: 297). In a note written on the end pages of his journal, Robinson carries on this thought and writes (*ibid*: 613):

> *"[T]he Aboriginal females on the islands have round circles cut in their flesh in imitation of the sun or the moon. I have seen a woman with four of them on her body; others I have seen with two or three. They are very fond of them, are generally placed on each side of the backbone and about the hips... The cicatrices of the sun and moon is intended to remove inflammation and having the power of those luminaries they imagine it will have the same influence on the part infected."*

Similar circular images are reproduced in rock engravings, drawings, huts, stone arrangements (Bonwick, 1870: 192), and on bodies, often beholding more than one meaning. Robinson writes of a surveyor, "Mr Hellyer", seeing a circular charcoal drawing and believing it was a representation of the sun. Robinson corrects him in his journal stating, "Those circles are emblematical devises of men and women" (ibid: 575). In regard to this entry, Plomley addresses the conflicting meanings without mentioning the possibility of the circle being a polysemous symbol. The moon was previously identified as a woman named Vetea, indicating a circle can mean both woman and moon. The multi-layered meanings of man, woman, moon, and sun are interchangeable and complex. The power of each is not confined to a singularity, but rather an Indigenous view of wellbeing, traversing body, environment, and spirit in an ebb and flow of meaning and balance.

Robinson identifies women specifically in the above passage, noting their cicatrices are localised around the hips and on either side of the spine. These areas on a woman's body are affected by strain during childbirth and menstruation. The waxing and waning moon is often linked to the cyclic flow of menstruation (Berndt and Berndt, 1993). The moon is recorded as both male and female across Aboriginal communities in Australia (usually male), and is often related to fertility, no matter the gender. The moon man in some traditions, if looked at directly can impregnate young women (Haynes, 1997: 107) or oppositely render the onlooker barren (Bates, 1972).

The placement of these cicatrices could be used as a healing agent in response to back pain and curing issues around fertility. Women were assigned much of the labour, including hunting crayfish, seals, climbing trees for possum, mining ochre, and on Robinson's journeys carrying the bulk of the load while travelling. The men hunted larger game and acted as guards for the group (Johnson et al., 2015; Robinson and Plomley, 2008).

Finally, the origin story recounted by Leigh Maynard (Thompson and Tasmanian Aboriginal Community, 2011) in Section 3.1 describes the phases of the moon. In the beginning, the Sun and Moon rose together (New Moon). As each day passed, the Moon woman fell behind the Sun man in his journey. He encouraged her by lighting more of her up each day, explaining the waxing moon. Eventually she was on the opposite side of the sky to the Sun (Full Moon). This is one of the rare accounts that explicitly acknowledges that the light of the Moon is a reflection of the Sun’s light (a point noted by R.S. Fuller, personal communication).

## 3.7 Transient Phenomena

Transient phenomena, such as meteors, eclipses, and aurorae, are featured prominently in Aboriginal traditions across Australia (Hamacher, 2012). Palawa from across Tasmania also have traditions of these phenomena, which are discussed in this section.

### 3.7.1 Meteors

There are few records of how Palawa perceived or understood meteors in their traditions. In southern Tasmania, a meteor is called *Pachareah* (Milligan, 1866: 426) and Coon (1972: 288) mentions that a falling meteorite at night startled some Palawa, who shrieked and hid their heads. In Plangermairrener traditions (Noonuccal, 1990: 115–119), a cheeky woman named *Puggareetya* tormented and fought a snake. Their wrestling upheaved the ground, forming the hills and mountains of the landscape. The snake cast the woman into the sky and is held there by the sky spirit *Mienteina*. *Puggareetya* continues to play tricks on the sky deities, who occasionally grow frustrated with her antics and throw her across the sky. She is then seen as a meteor (Hamacher and Norris, 2010).

As discussed in Section 3.1, the star-spirits, *Moinee* and *Droemerdeenne*, battled and *Moinee* fell to Earth at Cox Bight, where he can be seen today as a large standing stone (Coon 1972: 288). It is assumed *Moinee* took the symbolic form of a meteor (“falling star”), but this is inferred, never stated.

### 3.7.2 Aurorae

Cultural traditions of the Aurora Borealis (northern lights), which are commonly visible to cultures at high latitudes, tend to be associated with positive omens (Hamacher, 2013). Where aurorae are less common, such as those in the Southern Hemisphere, traditions err towards caution and act as warning. The positioning of Australia on the northern edge of the southern auroral zone means seeing the Aurora Australis is relatively rare, compared to areas within the peak of the auroral zones. Aurorae in Aboriginal traditions are often associated with blood, fire, and death because of its sometimes reddish appearance (*ibid*).

The Aurora Australis is well known to the Palawa of Tasmania, as the island lies at the upper edge of the southern auroral zone. There are a few different Palawa names of aurorae, as noted in Robinson's journals. On 19 October 1837, Robinson recorded two names from Rolepa, a leader of the Ben Lomond group, as *Nohoiner* and *Purnenyer*, and two names from the Western Palawa: *Genner* and *Nummergen*.

*Nohoiner* is nearly identical to the Cape Portland name *Noiheener*, attributed to an "electric spark" recorded in an entry by Robinson six years earlier. The Ben Lomond Palawa were thought to be linked in trade agreements with the Cape Portland Palawa. It is possible that they share language and it is possible the two words mean the same thing with respect to random light phenomena (Ryan, 2012: 32).

The earlier use of the Cape Portland word *Noiheener* was recorded by Robinson on 12 August 1832 and parallels the sentiments of mainland Aboriginal Australia's feelings of apprehension at an aurora (Robinson and Plomley, 2008: 430):

> *"The natives last night saw an electric spark in the atmosphere, at which they appeared frightened, and one of them told them not to mention it as they would all be sick if they did - the native of Cape Portland call in NOI.HEE.NER and the Port Sorell natives call it NAR.NO.BUN.NER.*"

It is unclear if the "electric spark" was referencing aurora. Similar words with slightly different spelling variations are applied to various forms of light phenomena, including aurorae, lightning, and thunder. *Nowhummer* is a word used by Aboriginal people from West Point and Cape Grim in Tasmania's Northwest of an evil spirit (Plomley, 2008: 650). People from Bruny Island are also recorded as believing thunder and lightning is an evil spirit (ibid: 321). In Plomley's consolidated word list, *Noiheenner* is a name given by various language groups to represent "God", good spirit, sun, moon, thunder, and lightning. These words may first appear to be different yet they all share attributes of ancestral deities. Robinson, being a religious man, may have translated meanings of thunder and lightning to God or spirits, all of which are taught to be respected and feared.

Records of auroral traditions in Tasmanian languages provide insight into how Palawa pay close attention to properties of natural phenomena. According to Anonymous (1877):

> *"There was a splendid Aurora in 1847, grand in its-effects at Hobart Town; and an interim one September 4, 1851, at the same place where the vividly shooting streamers of violet, red and other colors, where somewhat marred by the bright moonlight. The Aborigines of Tasmania compared the crackling noise of the curruscation to the snapping of their fingers."*

Despite reports of sound associated with aurora, it was not believed aurora could produce these sounds, as it was too far away. In 2012 researchers from Finland found a direct link to between noise and aurorae (Laine, 2012). They found that auroral sounds are actually born close to the ground.

### 3.7.3 Eclipses

There are no confirmed accounts of solar eclipses in recorded Palawa traditions, but there is a record of a lunar eclipse. During the Robinson expedition from 1829-1834, 11 lunar eclipses were visible from Tasmania, including two total eclipses (both in 1830)[2]. But only one was mentioned in any of Robinson's journals and none were identified from the remaining literature sources.

On 24 August 1831, Robinson writes that two days earlier, Manalargenna, Kickerterpoller, and three women left to make contact with other people in the area. They were away from Robinson's party for five days. During their absence, the guides with Robinson noticed the moon move into the Earth's shadow. They took this as an ominous sign that harm had come to Kickerterpoller and he had ascended to the moon. Truganini and Woorrady saw the lunar eclipse from Waterhouse point and read it as a bad sign that Robinson had been speared (Cameron, 2015). We identify this as a reference to a partial lunar eclipse visible on 23 August 1831 that reached mid-eclipse at 22:00. The perception of the eclipse by Truganini, Woorrady, and Robinson's guides is roughly consistent with other Aboriginal views of eclipses from across Australia (Hamacher and Norris, 2011).

## 4 SUMMARY AND CONCLUSION

This paper explores the fragments of Palawa astronomy recorded in the literature and archival documents dating back to the early nineteenth century, for which we attempt a partial reconstruction. While variations of knowledge in some cases are evident, there is continuity with many of the traditions, including those related to the Sun, Moon, the creation brothers, the stingray, calendars, time keeping, and views of transient phenomena. This suggest that Palawa use the sun for navigation and developing songlines.

Mainland Aboriginal traditions share fundamental similarities with those of Aboriginal Tasmanians. Locality affects individual groups' astronomical traditions across Australia, as the adaptive nature of the traditions reflects the natural world in which the community lives. Astronomical objects commonly associated with Aboriginal traditions on the mainland of Australia are the Milky Way, Orion, the Pleiades, the Magellanic Clouds, dark nebula (Coalsack), the Sun, and Moon. All are represented in recorded Tasmanian traditions except for the Pleiades and Magellanic Clouds. The absence of these objects is peculiar. They are incorporated into traditions of nearly all Aboriginal groups across Australia. Johnson (2011: 295) believes it is unlikely there are no Tasmanian traditions about the Pleiades, but for some reason they were simply never recorded.

This paper is a preliminary study into how Palawa construct and utilise the connection between the landscape and skyscape. This includes the diurnal motion of the sun and its application to navigation, how the movements of the stars were used to denote seasonal change and timekeeping, and how transient astronomical phenomena are associated with death or bad omens. The Moon's importance as a symbol of restoration and healing may have symbolic representation on cicatrising marks found on people's bodies and explained through oral traditions. This research shows how the night sky is a blackboard on which traditions are drawn with stars, retold to educate generations about moral code and law. But it is only a rudimentary starting point for future research.

## 5 Acknowledgements

We dedicate this paper to Aboriginal Tasmanians and elders, past and present. We thank Bob Fuller, James Goff, Carla Guedes, Trevor Leaman, Ben Silverstein, William Stevens, Rebe Taylor, and the the anonymous referees for comments and criticism.

Authors are listed in order of contribution to this paper. Lischick completed a research project on Tasmanian Aboriginal astronomy in Hamacher's third-year undergraduate course, *ATSI 3006: The Astronomy of Indigenous Australians*, at the University of New South Wales in early 2014. Gantevoort continued the research for a BA(Hons) thesis in Australian Indigenous Studies at UNSW, under the supervision of Hamacher. Hamacher worked closely with the students at all stages and completed his contributions as a staff member of Monash University.

Hamacher acknowledges support from the Australian Research Council (DE140101600).

## 6 Notes

1. https://en.wikipedia.org/wiki/Climate_of_Tasmania
2. Javascript Lunar Eclipse Explorer, NASA. Eclipse predictions by Fred Espenak and Chris O'Byrne. http://eclipse.gsfc.nasa.gov/JLEX/JLEX-index.html

## About the Authors

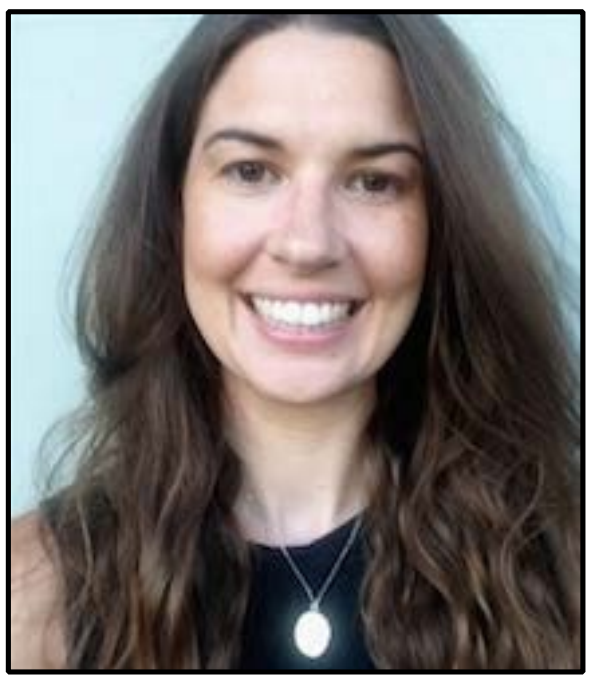

Michelle Gantevoort is an Operations Assistant at SBS Television in Sydney. She completed a B.A. in dance and theatre and a Master of Communication at UNSW. In 2015, she completed a B.A. with Honours in Indigenous Studies with a thesis on Tasmanian Aboriginal astronomy at UNSW, which was awarded First Class Honours. Michelle enrolled in a PhD program to study Indigenous Astronomy at Monash University in 2017 under the supervision of Dr Hamacher.

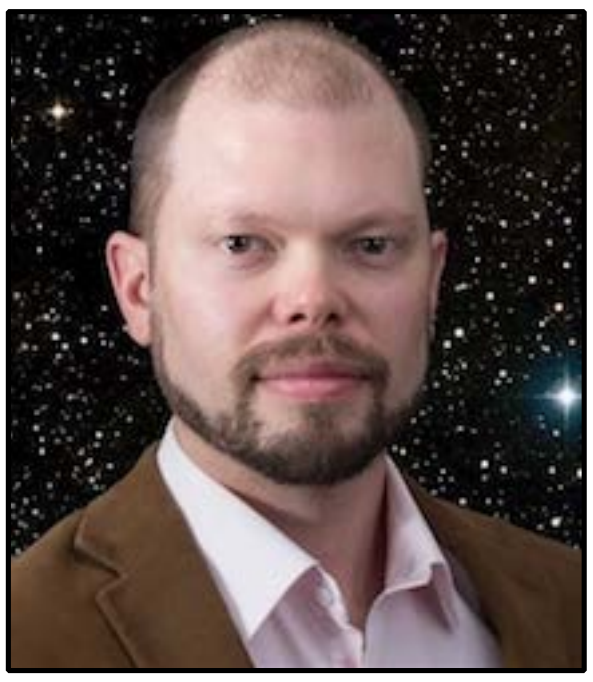

Dr Duane W. Hamacher is a Senior Australian Research Council (ARC) Discovery Early Career Research Fellow at the Monash Indigenous Centre in Melbourne. His research focuses on cultural astronomy with a focus on Australia and the Pacific. He earned a B.S. in physics, an M.S. in astrophysics, and a Ph.D. in Indigenous Studies. Duane serves as an Associate Editor of the *Journal of Astronomical History and Heritage*, is Secretary of the *International Society for Archaeoastronomy and Astronomy in Culture*, and Chair of the *International Astronomical Union* Working Group on Intangible Heritage.

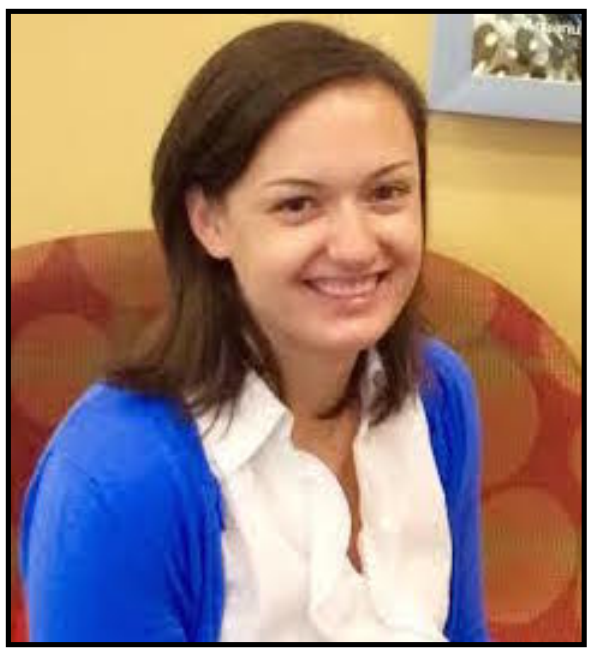

Savannah Lischick is a Process Engineer at LifeCell Corporation in Somerville, New Jersey, USA. She completed B.S. and M.S. degrees in Biomedical Engineering at the New Jersey Institute of Technology. Savannah spent a semester abroad at UNSW through the Global Engineering Program in 2014, where she enrolled in *ATSI 3006: The Astronomy of Indigenous Australians*, taught by Dr Hamacher. Lischick completed a research project on Tasmanian Aboriginal Astronomy, for which she received a High Distinction.